\documentclass{article}   
\usepackage{graphicx}				
\usepackage{amssymb}	          		
\title{A Proposal for an\\  ALPs-Chameleon Experiments Station - ACES\\}
\author{James R Boyce \\
\em{The College of William \& Mary and Jefferson Lab}  \\
Andrei Afanasev \\
\em{The George Washington University and Jefferson Lab}  \\
Oliver Keith Baker \\
\em{Yale University} \\
Michelle Shinn\\
\em{Jefferson Lab}
}
\date{March 24, 2014}						
\begin{document}
\maketitle
\section{Abstract}
It is generally accepted that certain astronomical and cosmological observations can be explained by invoking the concepts of Dark Matter and Dark Energy (DM/DE). Applying straightforward extensions of the Standard Model  to DM/DE, results in scalar fields and predictions of particles generation via photo-magnetic coupling . Under the right conditions, these particles should be observable in earth-bound laboratory settings. Although many attempts have been made to observe these particles, none have succeeded. 

Heretofore, most searches have focused on detecting multi-GeV Dark Matter WIMPS.  Recently, however, searches have been conducted in the lighter Dark Matter, sub-eV, WISP mass range. By comparison, little has been done to search for Dark Energy particles. The ALPs-Chameleon Experiments Stations (ACES) program, described herein, proposes a compact station that would search for both Dark Sector particles.\footnote{See table before References for acronyms used in this paper.}  

Finally, it is noted that both ''species'' of particles - dark energy and dark matter - could be generated at the same time in the same magnetic field with the possibility of interaction between DM and DE particles.  Thus, by using standard matter tools to produce particles from both dark sectors, ACES potentially could provide tri-sector discoveries with huge results for very little investment.

\section{Introduction}
Mysterious dark matter and dark energy have yet to be manifested in laboratory-based experiments despite predictions by theoretical models consistent with extensions of the Standard Model (SM). Many laboratory-based searches have focused on Axion and Axion-like particles or WISPs, (Weakly Interacting Sub-eV Particles).  Largely based on predictions of \cite{Jaeckel:2010ni}, a number of these experiments, including LIPSS at Jefferson Lab, \cite{Afanasev:2008jt}  \cite{Afanasev:2009fv}
used the LSW (Light Shining through a Wall) technique with the goal of detecting dark matter particles.

Until recently, dark energy manifestations, on the other hand,  have not been explored in earth bound laboratory settings. Theoretical treatment of dark energy predicts photo-magnetic coupling   particles with characteristics that depend on their environment, thus earning the name ''chameleon''.  \cite{Khoury:2003rn} This feature also means that chameleons formed in a vacuum chamber, are trapped in the vacuum chamber, until they re-enter the magnetic field and decay by the inverse process by which they were original generated. Detection of photons from such after-glow decay would be strong evidence of chameleon generation. 

The CHASE collaboration \cite{Chou:2008gr} conducted a series of afterglow experiments covering decay times from 120 seconds to several hours. However, due to a large ''orange glow'' background, CHASE was unable to explore afterglow decay rates shorter than 120 seconds. ACES proposes measurements of afterglow that covers the short time frame afterglow region and to verify CHASE results for longer time frames. 

The ACES Team proposes a three year effort to augment the LIPSS setup by adding a 10 W table top laser, a third dipole, an optical cavity (to increase the photon density by 1000x), and computer controlled shutters and insertion devices.  Then following testing and commissioning, and under automated computer controls, acquire and analyze after glow decay rates from 0.1 sec to 200 seconds. The first year would be devoted to construction and commissioning the system. The second and third years would be data acquisition, analysis and report writing. Final reports and results would be submitted for peer reviewed publication.

If nothing is found, the exclusion parameter space would be expanded. However, If evidence of either NLWCPs or chameleons are detected, it would open the door to new physics and perhaps the discovery of new forces. ACES  could produce game changing discoveries with huge results for very little investments.
\\
\\
\section{ACES Configuration} 
The proposed system is presented in three sub-systems. In so doing, the final full configuration better understood.
\subsection{ACES Configuration - Part 1: ALPs generation from left to right.}
\begin{figure}[htb]
\begin{center}
\includegraphics[width=\textwidth]{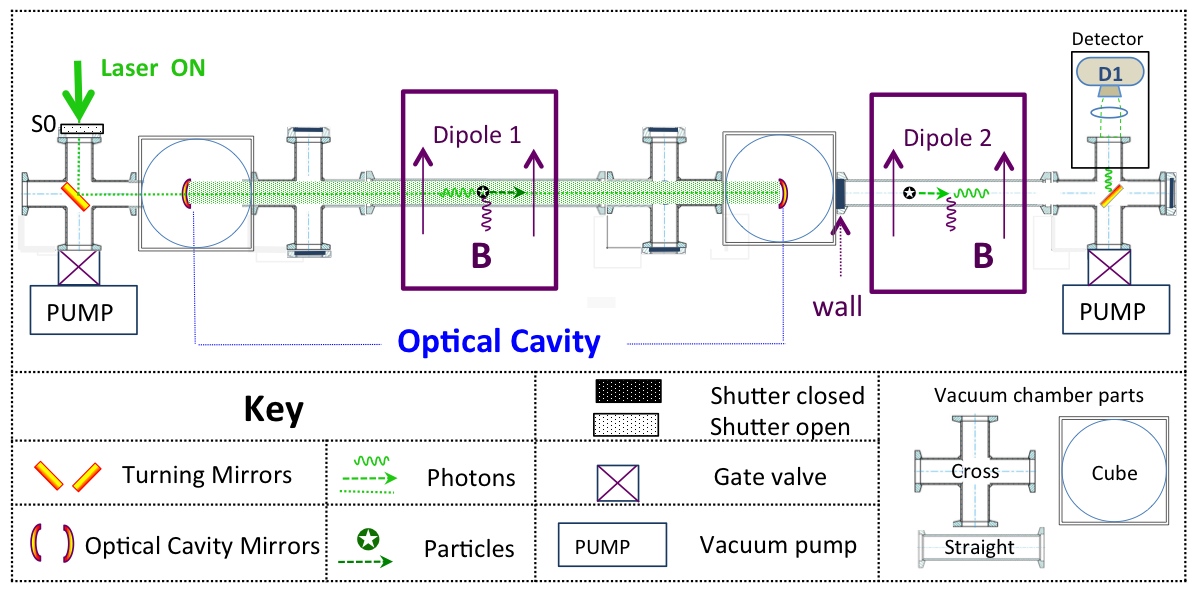}
\caption{ACES Configuration - Part 1: ALPs generation from left to right. Photons from a table-top laser, are stored in the optical cavity in a vacuum chamber inside the magnetic field of Dipole 1. According to theory, the dark matter scalar field can mediate photo-magnetic coupling resulting in ALPs and/or WISPS.
These dark matter particles, traveling in the same direction as the generating photon, pass through the wall, and enter the B-Field of Dipole 2.
The same coupling mechanism that generated the particles can re-generate them back into photons. The photons are then directed by the TM to Detector 1 where they are recorded. 
}
\label{ACES.layout.part1}
\end{center}
\end{figure}

The first sub-system is illustrated in Figure \ref{ACES.layout.part1}. The vacuum chamber will be assembled with conflate flanges to insure that it is light tight. The Detector, D1, is a Princeton Instruments LN$_2$~cooled CCD camera  specifically designed for low light conditions and is mounted in a light tight box with a focusing lens between the turning mirror and detector D1. Photo-magnetic coupling is illustrated inside the vacuum pipe inside the dipoles D1 and D2. A blank-off flange isolates the vacuum system into two sections. Each section is pumped down to high vacuum and valved off during data acquisition. Periodic checks will be conducted to insure vacuum integrity. The extra ports will be needed and described in the following layout sub-sections. 

\subsection{ACES Configuration - Part 2: ALPs generation from right to left.}

\begin{figure}[htb]
\begin{center}
\includegraphics[width=\textwidth]{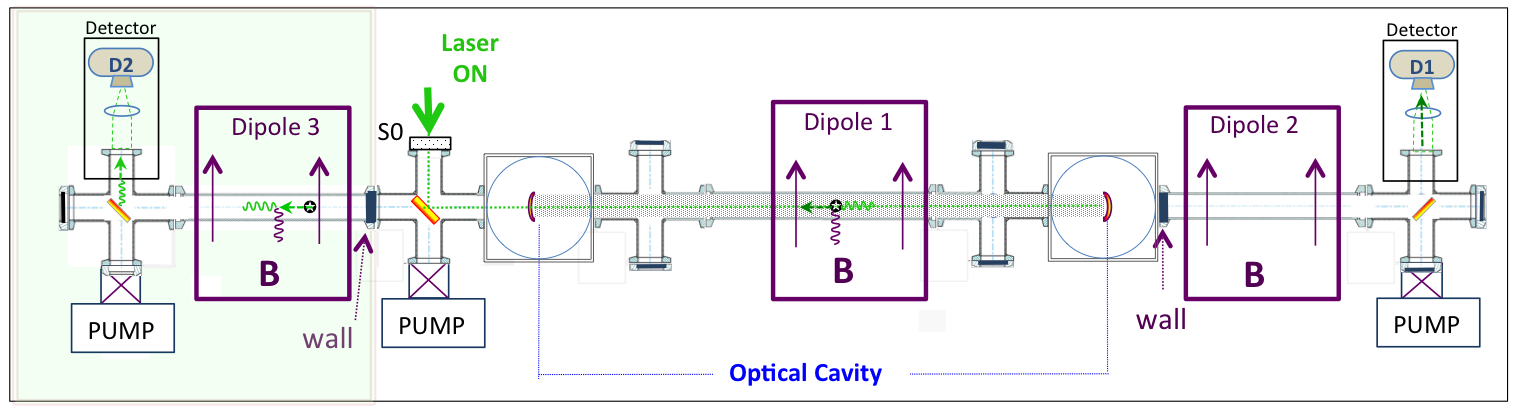}
\caption{ACES Configuration - Part 2: ALPs generation from right to left. 
Half of the photons stored in the optical cavity are traveling from right to left. By adding  a third dipole and detector - outlined in light green - ACES can capture photons regenerated from right-to-left traveling ALPs events otherwise lost.
}
\label{ACES.layout.part2}
\end{center}
\end{figure}

Since half  of the photons in the optical cavity are traveling from right to left, half of any dark mater particles generate by photo-magnetic coupling would be lost without a means to detect them. The addition of a regeneration section - a duplicate of the section just discussed - could collect  particles otherwise lost.  This is illustrated in Figure \ref{ACES.layout.part2}.  
\\
\subsection{ACES Configuration  - Part 3: Chameleons}

\begin{figure}[htb]
\begin{center}
\includegraphics[width=\textwidth]{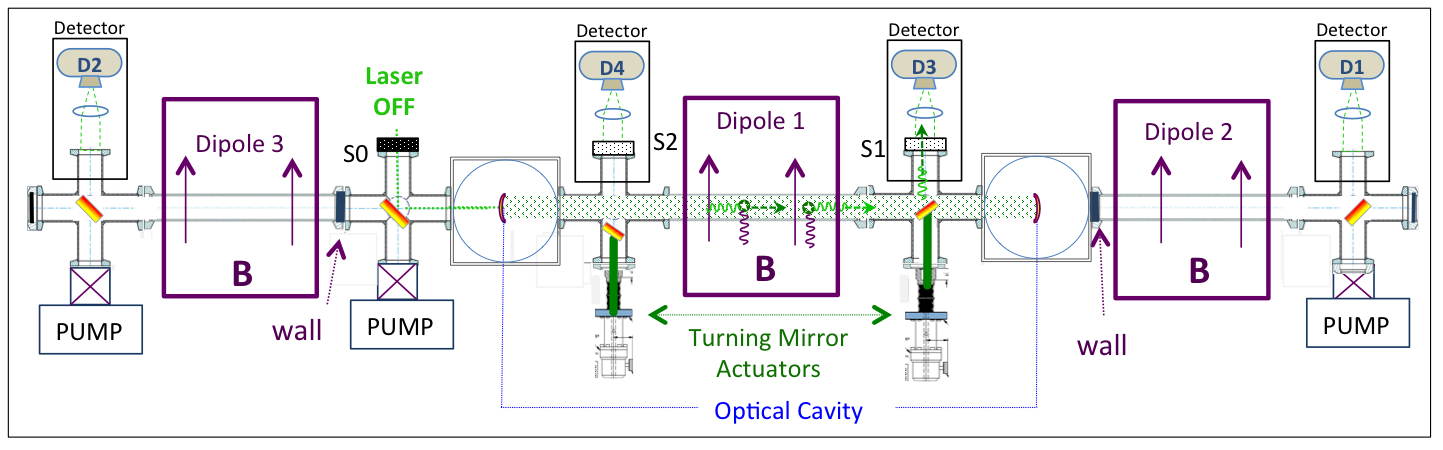}
\caption{ACES Configuration - Part 3: Chameleons. Theoretical treatment of Dark Energy  yields dark energy particles, dubbed ''chameleons''. Chameleons, unlike dark matter particles, cannot escape the confines of the vacuum chamber. Chameleons bounce around in the vacuum chamber like a gas, until, upon re-entering the B-field of Dipole 1, undergo coupling back to photons  traveling in the direction of the chameleon, either to the right or to the left. The intensity of the photon beam in the OC prevents direct measurement of decaying gas. However, once the photon source is turned off, photons from chameleon gas decay can be measured. 
}
\label{ACES.layout.part3}
\end{center}
\end{figure}

Theoretical treatment of Dark Energy predicts a scalar field that can mediate photo-magnetic coupling to produce dark energy particles, dubbed ''chameleons''. However, these dark energy particles, unlike dark matter particles, cannot escape the confines of the vacuum chamber. Generated chameleons would bounce around in vacuum chamber as a chameleon "gas".  Whenever a chameleon's travels re-enters the B-field of Dipole 1, it could undergo coupling back to photons traveling in the direction of the coupled chameleon, either to the right or to the left. Photons from chameleon decay, ''after-glow,''  can be measured with ACES by attaching two additional detectors and insertable mirrors on either side of Dipole 1, as is illustrated in Figure \ref{ACES.layout.part3}. 
\subsection{ACES Full Configuration}
\begin{figure}[htb]
\begin{center}
\includegraphics[width=\textwidth]{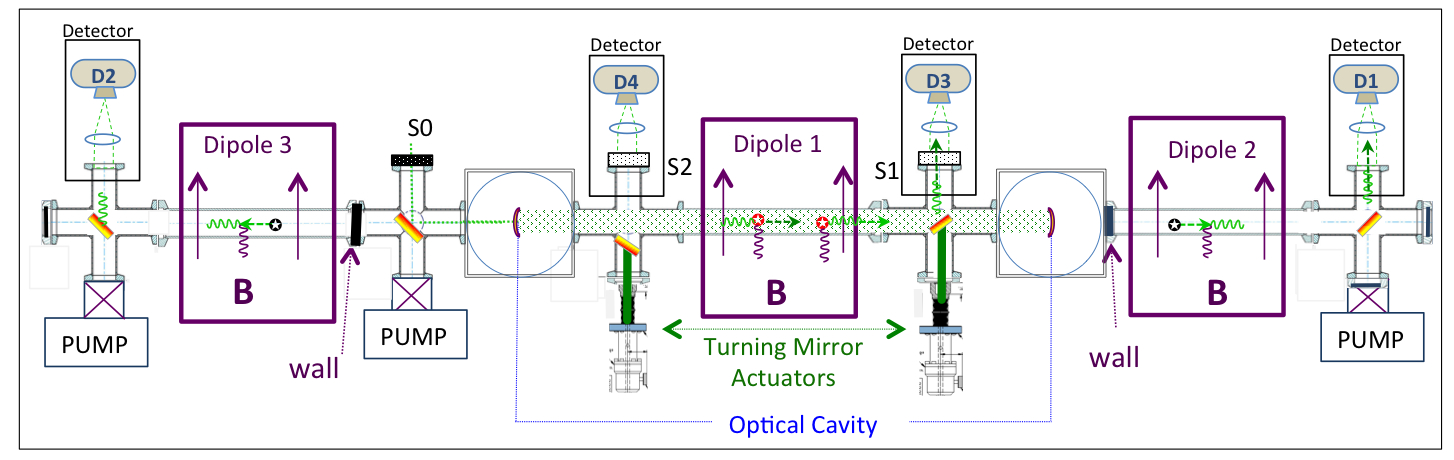}
\caption{ACES - Full Configuration. For illustration purposes the chameleon turning mirrors are shown in two positions: retracted for ALPs running and inserted for chameleon afterglow running. }
\label{ACES.Full.layout}
\end{center}
\end{figure}

The full configuration is illustrated in Figure \ref{ACES.Full.layout}. ALPs, if generated, are measured in Detectors D1 and D2. During ALPs measurements, actuators retract the chameleon turning mirrors  and the shutters S1 and S2 are closed and the chameleon gas is built up. Then, to measure the chameleon afterglow, the source laser is turned off, input shutter S0 is closed, actuators insert turning mirrors, shutters S1 and S2 are opened and after-glow decay is recorded. The procedure is repeated after retraction of the turning mirrors and shutters are reset.
Thus ACES is a system that can conduct simultaneous dark matter and dark energy experiments. 

The full configuration for ACES is shown in Figure \ref{ACES.Full.layout}. Starting with the LIPSS setup, \cite{Afanasev:2008jt} install a well cleaned vacuum system, an optical cavity - with mirrors specific to the wavelength of the laser light -, and actuator controlled mirrors. Also, install computer controlled shutters: S0, at the laser entrance window and S1 and S2  in front of detectors D3 and D4. S0 is open when the laser is on but closed when it is off. S1 and S2 are closed when the laser is on - to protect the cameras - and open when the laser is off and afterglow is being measured. The turning mirrors divert afterglow photons into Detectors D3 and D4. Ring-down time, the time for the laser light to decay away completely, is less than 40 microseconds. Thus by the time shutter S0 is closed, TMs inserted, and shutters S1 and S2 are opened, the only photons to be recorded by the camera are either from electronic noise, fluorescense, and/or chameleon decay, and the latter only if Dipole 1 is ON. \\

\section{Pre-Measurement Commissioning}
The vacuum system will be removed from the dipoles, rigorously cleaned, undergo vacuum bake-out, re-installed into dipole, hot nitrogen purged, and with an RGA, checked for residual hydrocarbon contaminations.  Detectors D1-D4 will be mounted in light-tight-boxes, attached to the vacuum system, and checked for light leaks. After installing the optical cavity,  a final 
hot N$_2$ purge will remove any residual hydrocarbons. Final alignment of the entire system will be conducted with a low power laser.  Once satisfactory alignment is achieved, it will be verified with the actual laser to be used:  a 10 W, green, CW table top laser. The cavity is expected to boost the photon density in Dipole 1 by  a factor of 1000, an amount routinely achieved with today's technology. Pitch, Yaw, and separation distance are all remotely controlled by pico-motors. Turning mirrors (TM)  will be mounted on the actuators and aligned. Software controlling the system will be tested and de-bugged. The system will be checked for any fluorescent light induced by the green laser. If any is found, it will be carefully measured as a background source and to determine if the fluorescence changes with dipole magnetic field. Finally, background measurements will be measured with Dipole field OFF, and with and without laser light.

\section{Experimental Procedure}
The measurement procedure is straightforward once the system is fully commissioned. With the Dipole 1 OFF, TM-Actuators retracted, and shutter S0 opened, the source laser is directed into the vacuum chamber. When the photons in the optical cavity have reached  the stored maximum level, ALPs data are recorded in Detectors D1 and D2. At the end of each ALPs run, simultaneously turn OFF the source laser, close shutter S0, insert TMs, and open shutters S1 and S2. Then record chameleon data in detectors D3 and D4 for a preset time interval. At the end of the interval, close shutters S1 and S2, retract TMs, open S0 and  turn ON the source laser. These steps are then repeated until statistically significant data are accumulated. The procedure will be computer controlled. ALPs and afterglow afterglow data runs are measured a second time with the dipole B-field ON. Significant variations in the two sets of data (dipoles ON and OFF)  will be repeated and examined. Initially, time intervals  0.1, 1, 2, 5, 10, 15, 25, 50, 75, 10, 150, 200, and 300 seconds are planned. Other intervals may be added if needed.   

\section{Discussion}
Vital to the success of any search such as that proposed  by ACES is a thorough treatment of backgrounds. In the case of CHASE, the large background characterized as ''orange-glow'' has been attributed to photoluminescence induced by the green laser on small amounts of organic material, e.g., vacuum grease. Hence ACES emphasis on a well cleaned vacuum chamber.

Another possible source of background is laser induced Photoluminescence (PL) from the stainless steel of the vacuum chamber...at low temperature: $4\,^{\circ}\mathrm{K}$. 
At non-cryogenic temperatures, metals do not exhibit PL since there are several non-radiative paths back to the ground state, paths such as vibrational or internal conversion. However, as the temperature of the metal is lowered, these non-radiative paths become fewer and fewer, ultimately leaving only the radiative paths available for de-excitation.  \cite{0957-0233-20-7-075304}.

It is speculation at the moment, but the source of CHASES ''orange glow,'' could be due to PL from the stainless steel vacuum chamber of the Tevatron dipole, which was at $4\,^{\circ}\mathrm{K}$ during the experiment. The green laser used for chameleon generating photo-magnetic coupling, could induce PL in the orange portion of the visible spectrum, thus contributing to the observed high background. A separate research effort to explore this possibility is now underway with the SRF Institute at Jefferson Lab. If low temperature PL is relatively strong, then PL background will not be an issue with ACES but could explain the high background that plagued the CHASE effort.
Other sources of background are electronic noise, cosmic rays, and light leakage into the system. All of which are understood. The experience gained by the LIPSS Collaboration will be put to good use.

\subsection{Speculation: DM/DE interaction.}
Now that ACES is described, we realize that when the system is fully operational, both dark matter and dark energy particles could be present in the vacuum system and could interact. This interaction has not been examined, to our knowledge. If the probability  of generating and regenerating an ALPs is P$_A$ and the probability of generating/re-generating a chameleon is P$_C$, the detection of and APS is P$_A$*P$_A$ and for chameleons P$_C$*P$_C$.  The probability for interaction of an ALPs with a Chameleon would be P$_A$*P$_C$ times the cross section for interaction. What that interaction would produce is an interesting question for theorists. If it results in  a standard model particle, it might be possible to modify ACES to look for these cross-sector events.

\section{Conclusions}
A case has been made for setting up and running an ALPs - Chameleon Experiments Station for DM/DE searches at Jefferson Lab. The LIPSS set-up in Lab 1 of the FEL, can be augmented with the addition of an optical cavity -  to boost a 10 W CW green laser by a factor of 1000 - a third dipole, three more detectors, two insertable turning mirrors, and  computer controlled data acquisition system. 
 
 ALPs results will be cross checked with previous measurements. 
Decay times for chameleon afterglow can be extended from the 120 second limit of CHASE to as short as 0.1 seconds, perhaps even lower. 

 Results and experience with the proposed system can provide valuable information for designing, funding, and running future laboratory-based DE/DM experiments, potentially including measurements of DM particles interacting with DE particles. 
\\
\section*{Acknowledgments} 
This unfunded proposal was developed under the auspices of the Jefferson Lab User program in partnership with the Office of Sponsored Programs at the College of William and Mary and the Applied Research Center.
\\
\section*{Acronyms}
\small
\vspace{0.2cm} 
{\renewcommand{\arraystretch}{1.5}
\renewcommand{\tabcolsep}{0.2cm}
\begin{table}[h]
\centering
\begin{tabular}{| l | l | l | l |}
\hline
\small
ACES & ALPs-Chameleon Experiments Station & ALPs & Axion-Like Particles \\
\hline
\small
CHASE & CHAmeleon Search Experiment & DM/DE & Dark Matter/Dark Energy \\
\hline
\small
LIPSS & LIght Pseudoscalar and Scalar Search & LN$_2$ & Liquid Nitrogen \\
\hline
\small
	NLWCP & New, Light, Weakly-Coupled Particles & OC & Optical Cavity \\
\hline
\small
WISPs & Weakly Interacting Sub-eV Particles & PL & Photo-Luminescence  \\
\hline
\small
WIMPs & Weakly Interacting Massive Particles & SM & Standard Model \\
\hline
\end{tabular}
\end{table} 
}

\end{document}